\documentclass[journal=apchd5,manuscript=perspective,layout=twocolumn]{achemso}
{}
\makeatletter \let\l@addto@macro\relax \makeatother
\usepackage[fontsize=11pt]{scrextend}
\setkeys{acs}{articletitle=true}

\usepackage[utf8]{inputenc}
\usepackage[T1]{fontenc}

\usepackage{graphicx}
\usepackage{amsmath,amssymb}
\usepackage[dvipsnames]{xcolor}
\usepackage[numbers,sort&compress,super]{natbib}
\usepackage[colorlinks,allcolors=blue]{hyperref}

\renewcommand{\Im}{\operatorname{Im}}
\newcommand{\rr}{\mathbf{r}}
\newcommand{\GG}{\mathbf{G}(\rr,\rr^\prime,\omega)}

\title[Molecular polaritonics]{A theoretical perspective on molecular polaritonics}

\author{Mónica Sánchez-Barquilla}
\affiliation{Departamento de Física Teórica de la Materia Condensada and Condensed Matter Physics Center (IFIMAC), Universidad Autónoma de Madrid, E-28049 Madrid, Spain}

\author{Antonio I. Fernández-Domínguez}
\affiliation{Departamento de Física Teórica de la Materia Condensada and Condensed Matter Physics Center (IFIMAC), Universidad Autónoma de Madrid, E-28049 Madrid, Spain}

\author{Johannes Feist}
\email{johannes.feist@uam.es}
\affiliation{Departamento de Física Teórica de la Materia Condensada and Condensed Matter Physics Center (IFIMAC), Universidad Autónoma de Madrid, E-28049 Madrid, Spain}

\author{Francisco J. García-Vidal}
\email{fj.garcia@uam.es}
\affiliation{Departamento de Física Teórica de la Materia Condensada and Condensed Matter Physics Center (IFIMAC), Universidad Autónoma de Madrid, E-28049 Madrid, Spain}
\alsoaffiliation{Institute of High Performance Computing, Agency for Science, Technology, and Research (A*STAR), Connexis, 138632 Singapore}

\begin{document}

\begin{abstract}
  In the last decade, much theoretical research has focused on studying the
  strong coupling between organic molecules (or quantum emitters, in general)
  and light modes. The description and prediction of polaritonic phenomena
  emerging in this light-matter interaction regime have proven to be difficult
  tasks. The challenge originates from the enormous number of degrees of freedom
  that need to be taken into account, both in the organic molecules and in their
  photonic environment. On the one hand, the accurate treatment of the
  vibrational spectrum of the former is key, and simplified quantum models are
  not valid in many cases. On the other hand, most photonic setups have complex
  geometric and material characteristics, with the result that photon fields
  corresponding to more than just a single electromagnetic mode contribute to
  the light-matter interaction in these platforms. Moreover, loss and
  dissipation, in the form of absorption or radiation, must also be included in
  the theoretical description of polaritons. Here, we review and offer our own
  perspective on some of the work recently done in the modelling of interacting
  molecular and optical states with increasing complexity.
\end{abstract}

\maketitle

\section{Introduction}

Polariton is a general term used to describe a hybrid light-matter excitation,
and has been employed in many different situations in the history of
physics~\cite{Basov2021}. In this Perspective, we focus on a small subset of
these situations, namely those in which molecules (often organic dyes) are used
to provide the material component. Even within this small subset, a wide range
of new phenomena are enabled by polariton formation, among them Bose-Einstein
condensation and polariton lasing~\cite{Kasprzak2006, Kena-Cohen2010}, quantum
information processing~\cite{Hennessy2007}, long-range excitation
transport~\cite{Coles2014, Zhong2017} and control of chemical reaction
rates~\cite{Hutchison2012}. After a short overview of basic physical concepts,
we discuss our view on the current state of the field and the challenges it
faces, interesting recent developments, and promising future directions. Despite
the focus on molecules, we here restrict the discussion mostly to ``physical''
properties, and ignore ``chemical'' properties such as (photo)reactivity that
have been the focus of a recent related perspective on polaritonic
chemistry~\cite{Fregoni2021Perspective}.

Polaritons arise when the interaction strength between
light modes and material excitations in a system becomes large
enough that the eigenstates of the system are not even
approximately represented by pure matter excitations or pure
electromagnetic excitations (photons), and instead become mixed. 
Polariton formation is thus
intimately related to the regime of strong coupling (SC) between
light and matter~\cite{Kimble1998,Torma2015}, reached when the
interaction strength overcomes the loss rates of the constituents.
We note that strong coupling itself is not necessarily a quantum
effect, and can often be modelled through classical
electromagnetism (EM)~\cite{Tanji-Suzuki2011Interaction,
Torma2015}. In that case, the material excitations are represented
through the dielectric function of the medium (e.g., through a
resonance of Drude-Lorentz form), or through a polarizable dipole
for single emitters~\cite{Kewes2018,SaezBlazquez2020}.

\begin{figure}[htbp]
  \includegraphics[width=\linewidth]{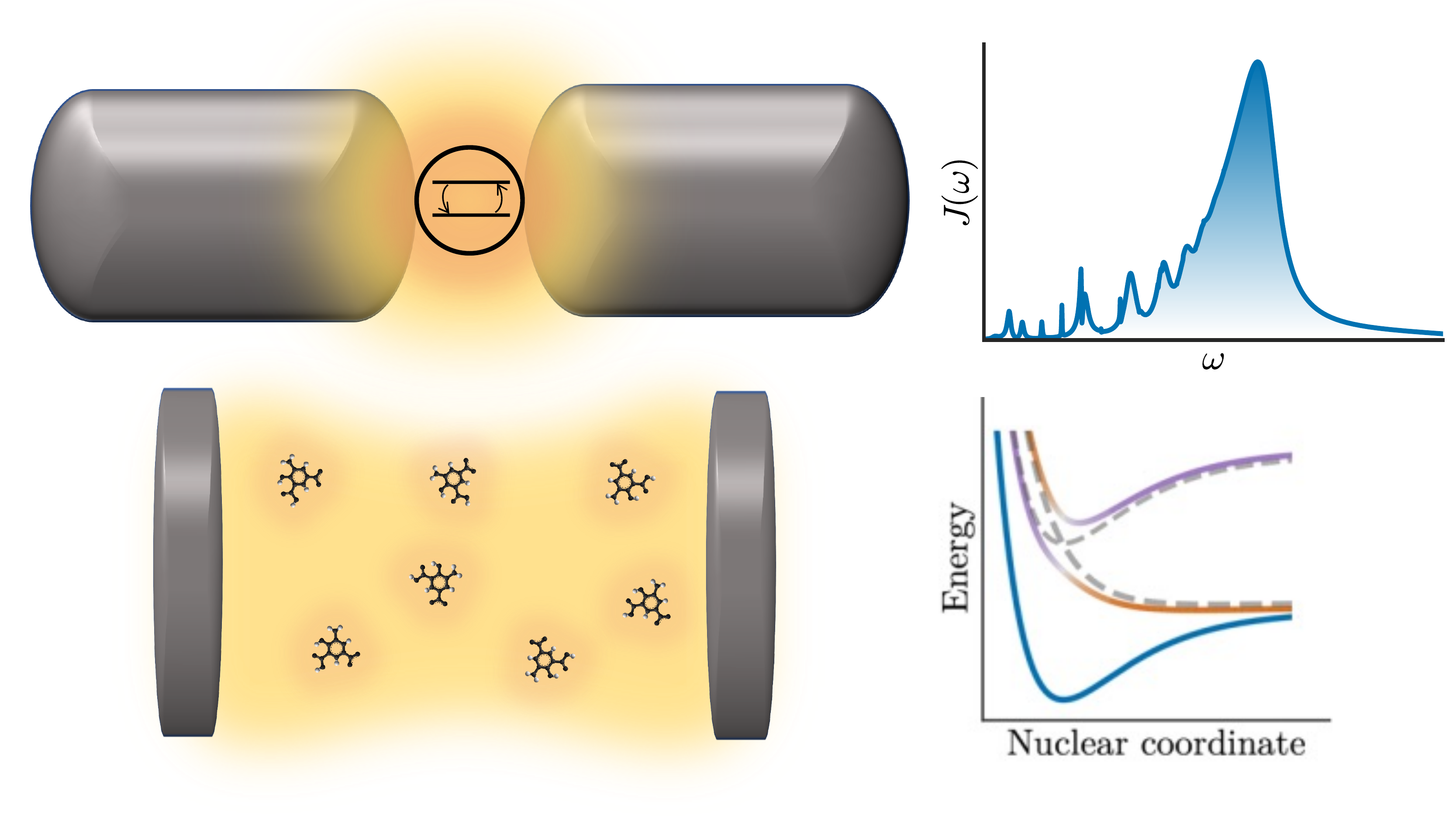}
  \caption{Schematic representations of typical situations in molecular
  polaritonics. Left: Single emitter coupled to a plasmonic nanocavity (top),
  and molecular ensemble interacting with a photonic microcavity (bottom).
  Right: Photonic spectral density (top) and polaritonic (red and violet)
  potential energy surfaces (bottom).}
  \label{fig:scheme}
\end{figure}

Before introducing strong coupling in more detail, we note that the apparently
simple question of what should be called a ``light mode'' is actually somewhat
subtle, and, as for any question of semantics, its answer is to some degree
arbitrary. The only ``pure'' light modes are free-space modes in vacuum. A
cavity is a material structure that confines the EM field, such that the
confined ``cavity photon modes'' are always partially material
excitations~\cite{Feist2020, Fregoni2021Perspective}. As an operational
definition, the cavity modes are usually understood as the modes supported by
those parts of the full system for which the relevant dynamics can be
well-approximated through macroscopic electromagnetism under linear response.
This distinction leads naturally to the framework of macroscopic
QED~\cite{Fano1956, Huttner1992, Scheel1998, Knoll2001, Scheel2008,
Buhmann2012I, Buhmann2012II, Feist2020}, discussed in more detail below. The
resulting EM modes then include those of optical (Fabry-Pérot)
cavities~\cite{Gerard1996,Hood2000}
(as depicted in the left bottom panel of \autoref{fig:scheme}), photonic
crystals~\cite{Painter1999,Lodahl2015},
and often also plasmonic nanostructures 
(as illustrated in the left top panel of \autoref{fig:scheme}), 
whose resonances are formed due to
geometrical restriction of the free electron motion in metals and allow strongly
subwavelength field confinement to be
achieved~\cite{Fernandez-Dominguez2017,Fernandez-Dominguez2018,Baumberg2019}.
However,
plasmon modes are physically quite distinct from ``optical'' modes. In
Fabry-Pérot or photonic crystal cavities, the energy is mostly stored in the
electric and magnetic fields, and the dielectric functions of the materials
can often be treated as approximately constant within the relevant range of
frequencies. In contrast, the energy in plasmonic resonances is stored in the
electric field and the kinetic energy of the electrons~\cite{Khurgin2015},
such that the resulting modes are more correctly referred to as surface plasmon
polaritons. The dominantly electrostatic (or more precisely, quasistatic)
character of the fields in deep-subwavelength cavities also has fundamental
consequences on the light-matter interaction, as the electric fields are then
to a good approximation purely longitudinal and not
represented by the vector
potential in the standard Coulomb gauge~\cite{Galego2019, Feist2020, Fregoni2021Perspective}.
Furthermore, plasmonic systems can have complicated internal dynamics after
excitation, e.g., leading to hot electron generation which in turn can have
significant effects on chemistry~\cite{Zhang2018Surface, Zhou2018Quantifying}.
In this Perspective, we do not discuss such hot-electron effects.

Once a suitable definition of what constitutes the light modes in the system has
been chosen, it becomes possible to distinguish between the weak and strong
coupling regime within that model. For a single quantum emitter approximated as a
two-level system coupled to a single photon mode within the rotating wave
approximation, the dynamics is described by the Jaynes-Cummings
model~\cite{Jaynes1963}. When the emitter and photon mode are on resonance, the
strong coupling regime is entered when their mutual interaction overcomes
decoherence in the system. In this
regime, energy exchange between light and matter becomes a coherent process: if
only one of the components is initially excited, this energy exchange is seen as
an oscillatory behaviour of the population between both subsystems, the so-called
vacuum Rabi oscillations occurring at the vacuum Rabi frequency (or Rabi
splitting)
\begin{equation}
  \Omega_R = 2g = \frac{2}{\hbar} \vec{\mu}\cdot \vec{E}_{1ph},\label{Rabi}
\end{equation}
where $\vec{\mu}$ is the dipole transition moment of the emitter and
$|\vec{E}_{1ph}| = \sqrt{\hbar\omega_c/2\epsilon_0V}$ is the quantized electric
field strength of the mode, associated with one photon, at the emitter position.
Here, $\omega_c$ is the cavity mode frequency and $V$ is its effective mode
volume, whose definition in  nanophotonic devices has been the object of intense
theoretical activity lately~\cite{Kristensen2012,Cognee2019,Tserkezis2020}.
Physically, vacuum Rabi oscillations thus correspond to the situation where a
photon can be emitted and reabsorbed several times before it disappears from the
system. In these conditions, the eigenstates of the coupled system are hybrid
light-matter states, and a thorough understanding of the system can only be
reached by considering the coupled system as a whole. The two original excited
states (emitter and photon) transform into polaritonic states that are shifted
up and down in frequency by the coupling strength, with their difference in
energy given by the Rabi splitting. These two states are conventionally called
the lower polariton (LP) and upper polariton (UP).

We mention here that some care should be taken to distinguish between the
concepts of polaritonic states and polaritons. The former are the hybrid
eigenstates of the coupled system, which do not depend on the state of the
system at any point in time and in that sense always exist. The latter are the
excitations of the system in a quasi-particle picture, and thus only exist when
the system is in one of the polaritonic states. As a further complication, for
(approximately) linear systems that are well-modeled as harmonic oscillators,
polaritons behave approximately like bosons. It is then convenient to use the
concept of (bosonic) polaritonic modes, each of which describes a (formally)
infinite number of polaritonic Fock states $|0\rangle_P, |1\rangle_P,
|2\rangle_P, \ldots$. In this picture, having $n$ polaritons in mode $P$ means
that the system is in state $|n\rangle_P$. In the literature, the distinction
between polaritons, polaritonic states, and polaritonic modes is not always made
explicit, which can cause some confusion. Which of the three is meant is usually
clear from the context.

For single emitters, reaching the SC regime is extremely challenging, 
as the single-emitter coupling
strength $g$ has to become comparable to the emitter and cavity decoherence
rates. This can be achieved by either increasing the coupling strength or decreasing the
decoherence rates sufficiently. This was first achieved in 1985 by working with
long-lived emitters and cavities at cryogenic temperatures, in a microwave
cavity with superconducting mirrors~\cite{Meschede1985}. Rabi oscillations in
the system were explicitly measured two years later~\cite{Rempe1987}. In 2004,
the SC regime for a single quantum dot in a semiconductor micropillar cavity was
achieved, with a Rabi splitting $\sim 100~\mu$eV~\cite{Reithmaier2004}. These
approaches, where the absolute coupling strength is a small fraction of the
excitation energy, necessarily require very long-lived emitters and cavity
modes, which in turn implies cryogenic temperatures. More recently, the strong
coupling regime has been approached at room temperature for single organic molecules by
using extremely localized surface plasmons in narrow
gaps~\cite{Chikkaraddy2016}, which support light confinement in deeply
subwavelength volumes~\cite{Santhosh2016, Chikkaraddy2016, Liu2017Strong,
Li2021Bright, Wu2021}. The estimated single-molecule Rabi splitting achieved in
Ref.~\citenum{Chikkaraddy2016} was $90$~meV, at the limit of the strong coupling
regime. Very recently, strong coupling and quantum nonlinearity have been
observed for a single molecule at cryogenic temperature~\cite{Pscherer2021},
where the molecule behaves as an effective two-level system.

Strong light-matter coupling is much easier to achieve in the collective case where
an ensemble of $N$ close to identical quantum emitters interacts with a photonic
mode (as described by the Tavis-Cummings model~\cite{Tavis1968}). In that case,
the effective coupling strength increases with the number of emitters as $g_N =
g\sqrt{N}$~\cite{Garraway2011}. This enhancement significantly simplifies
entering the strong coupling regime and is the basis for most experiments in the
field of molecular polaritonics. It occurs because an excitation in this case
can be coherently distributed over the $N$ emitters, forming a so-called bright
state with increased light-matter coupling. At the same time, all $N-1$
orthogonal ways of distributing an excitation over the emitters show negligible
coupling to the cavity mode due to destructive interference between the dipole
transitions in the different emitters. These superpositions are the so-called
dark states (DS), which play a major role in molecular polariton
dynamics~\cite{Feist2018, Ribeiro2018Polariton}.

Collective strong coupling was first realized in 1975 using molecular vibrations
coupled to surface phonon polariton modes~\cite{Yakovlev1975}, and soon later
for molecular excitons coupled to surface plasmon
polaritons~\cite{Pockrand1982}, and Rydberg atoms coupled to a high-Q microwave
cavity~\cite{Kaluzny1983}. Strong coupling to semiconductor (Wannier) excitons
was first realized in 1992~\cite{Thompson1992}. Such systems reach Rabi
splittings in the range of $1-20$~meV~\cite{Weisbuch1992, Houdre1994,
Kelkar1995}. Organic semiconductors support much larger Rabi splittings,
$\Omega_R \gtrsim 100$~meV, due to their high density and large dipole moments,
so that strong coupling can be observed at room temperature~\cite{Lidzey1998}.
We note that the maximally reachable Rabi splitting for a given material is
determined by the density of dipoles, but largely independent of the specifics
of the photon mode~\cite{Abujetas2019, *Abujetas2019Erratum, Canales2021,
Barra-Burillo2021}. This can be understood by noticing that
\begin{equation}
  g_N \propto \mu \sqrt{N/V} \propto \mu \sqrt{\rho},
\end{equation}
where $\rho$ is the molecular number density and we have used that the effective
mode volume is related to the physical volume occupied by the mode. The number
of molecules interacting with the mode is thus proportional to the molecular
density times the volume. A more careful calculation shows that the Rabi
splitting depends on the dipole density multiplied by a ``filling factor''
between $0$ and $1$ that determines what fraction of the mode volume is filled
with the molecular material (weighted with the position-dependent quantized
field strength)~\cite{Abujetas2019, *Abujetas2019Erratum,Tserkezis2020}.
When a cavity is
completely filled with the material in question, the Rabi splitting is equal to
the bulk polariton splitting obtained by Hopfield in 1958~\cite{Hopfield1958}.
These facts explain why similar Rabi splittings have been observed in the
literature for very different photonic systems, such as Fabry-Pérot
cavities~\cite{Lidzey1998}, plasmonic surfaces~\cite{Bellessa2004}, plasmonic
hole arrays~\cite{Dintinger2005}, isolated particles and arrays of
them~\cite{Zengin2015, Rodriguez2013}, and nanoparticle-on-mirror
setups~\cite{Chikkaraddy2016}. Several kinds of organic materials can reach the
ultrastrong coupling regime~\cite{FriskKockum2019}, in which the Rabi splitting
is a significant fraction of the bare excitation energy, with record values
close to and above $\Omega_R = 1$~eV~\cite{Schwartz2011, Gambino2014,
Eizner2018}. These large values also mean that cavity modes with very large
decay rates $\kappa$ (or equivalently, short lifetimes $\tau = 1/\kappa$ or low
quality factors $\omega_c/\kappa$) can be used while still reaching the strong
coupling regime.

Organic molecules are very well-suited for reaching large Rabi
splittings due to the large transition dipole moments and high densities.
However, they have complex internal structure due to their rovibrational degrees
of freedom and often cannot be approximated as two-level systems. On the one
hand, this complicates their use and study as idealized (two-level) quantum emitters. On
the positive side, this opens up the opportunity to modify their internal structure
and dynamics through strong light-matter coupling, or conversely to exploit the
internal dynamics to achieve new photonic functionalities. The former type of
applications are exemplified by the field of polaritonic chemistry, which aims
at modifying chemical processes such as photochemical reactions through strong
light-matter coupling~\cite{Fregoni2021Perspective}. The latter type of
applications typically rely on the fact that molecules show strong
exciton-vibration interactions, such that molecular vibrations can drive
polariton relaxation or transfer between different polaritonic
states~\cite{Litinskaya2004, Coles2011Vibrationally}. This can enable processes
such as organic exciton-polariton lasing and condensation~\cite{Kena-Cohen2010,
Ramezani2017Plasmon, Keeling2020} or energy transfer between different molecular
species even over long spatial distances~\cite{Coles2014, Zhong2017,
Garcia-Vidal2017, Du2018Theory, Saez-Blazquez2018Organic, Georgiou2021,
Satapathy2021}.
Note that it has been also shown that, even under the two-level system
approximation, SC phenomena involving organic molecules offer possibilities
for nonclassical light generation not attainable by means of other types of
quantum emitters~\cite{Saez-Blazquez2017,Saez-Blazquez2018Photon}.

When describing light-matter interactions in molecular systems, in particular in the strong coupling
regime, including all the degrees of freedom in both constituents is an arduous
task. Then, many models are focused on taking into account the complexity of one
of them, i.e., the theoretical effort is focused either on the description of the
complexity of the photonic structures or to include to some extent
the vibrational structure of the molecules. In what follows, we summarize some of the
theoretical challenges that remain in both paths.

\section{Complex EM fields under strong coupling}

Many different kinds of ``cavities'' can be used to achieve strong coupling in
the collective regime, while few- or single-molecule strong coupling necessarily
requires deep subwavelength confinement of light. In general, any cavity setup
is determined by ``macroscopic'' structures consisting of large numbers of
atoms, such as Fabry-Pérot cavities, photonic crystals, and metallic
nanoparticles or surfaces. Within these setups, one or several microscopic
quantum emitters such as atoms, molecules or point defects are placed. As
discussed above, it is then customary to treat the macroscopic structure through
Maxwell's equations and formally treat the modes arising from these equations as
the EM modes of the system.

In order to describe light-matter interactions on a quantum level, these EM
modes have to be quantized, which is significantly more challenging than the
quantization of free EM modes in conventional quantum electrodynamics. For
one, the presence of material structures complicates the solution of eigenmodes,
which often is only possible numerically. Nowadays, many commercial and open
source packages are available to solve Maxwell's equations.
Furthermore, these light modes will be lossy, often highly so, both due
to material losses and leakage to the far field. Only in some approximations do
lossless states exist, e.g., when assuming the existence of perfectly conducting
mirrors or infinitely long lossless and defect-free waveguides, etc. None of
these are typically good approximations for the kinds of structures used in
molecular polaritonics. Still, when losses are small enough, it can be a
reasonable strategy to quantize fully bound modes in a fictitious lossless
system, and then treat the losses as small perturbations on top of that.

Alternatively, when dealing with small enough nanoparticles with localized
resonances (such as plasmonic nanoparticles) for which radiative losses are
small due to inefficient emission, the so-called quasistatic approximation is
often applicable. In this approximation, retardation effects, and therefore EM
propagation into free-space (or bulk dielectric media) is neglected, resulting
in purely longitudinal fields. In this limit, semi-analytical solutions are
often again possible, e.g., using transformation
optics~\cite{Li2016Transformation, Li2018Plasmon, Huidobro2020CRP}, with the
resulting modes being fully bound while still describing the material losses.
Since subwavelength confinement is a prerequisite for the quasistatic
approximation, these material losses will always be
significant~\cite{Khurgin2015}. One advantage of the quasistatic approximation
is that the EM modes can be described by a scalar potential, which simplifies
the treatment of beyond-dipole interactions~\cite{Cuartero-Gonzalez2018,
Neuman2018Coupling, Fregoni2021Strong}. Another advantage of the quasistatic
approximation is that for metals described by a dielectric function of Drude
form, the resulting eigenmodes will always correspond to uncoupled Lorentzians
in the spectral density (discussed in more detail below), which allows for a
straightforward quantization procedure of the resulting
modes~\cite{Cuartero-Gonzalez2018, Cuartero-Gonzalez2021Distortion,
Fregoni2021Strong}. Radiative losses can also be included a posteriori, e.g., by
calculating the effective dipole moment of the localized
resonances~\cite{Cuartero-Gonzalez2018,Cuartero-Gonzalez2020Dipolar}.

When the quasistatic approximation is not appropriate and retardation effects
have to be taken into account, the most general and powerful approach to
nonetheless obtain a quantized description of the EM field is macroscopic QED\@.
This is a formalism that quantizes the EM field in arbitrary structures,
including dispersive and absorbing materials. A particularly appealing feature
of this approach is that the quantized EM modes are fully described by the
dyadic Green's function $\GG$ of the classical Maxwell equations, which can be
obtained from any numerical solver. This can be conceptually understood from the
fact that Maxwell's equations are the wave equations describing the dynamics of
EM fields, which remains true after quantization. A recent review about
macroscopic QED in the context of nanophotonics can be found in
Ref.~\citenum{Feist2020}. We note here that, although the classical description of
the EM environment is valid for a wide variety of physical situations, it breaks
down when the material and the emitters are physically close enough that
electronic wave functions overlap~\cite{Kulkarni2015,Babaze2021}, which happens
at subnanometer separations.

For an emitter where a single dipole transition is relevant (i.e., when treating
only two electronic levels), the so-called spectral density $J(\omega)$
completely characterizes the electromagnetic environment and its interaction
with the emitter (a schematic picture of this physical 
magnitude is rendered in the top right panel of \autoref{fig:scheme}). This quantity is given by
\begin{equation}\label{eq:spectral_density}
  J(\omega) = \frac{\hbar\omega^2}{\pi\epsilon_0 c^2} \boldsymbol{\mu} \cdot \Im\mathbf{G}(\rr,\rr,\omega) \cdot \boldsymbol{\mu},
\end{equation}
where $\boldsymbol{\mu}$ is the transition dipole and $\rr$ is the position of
the emitter. For the many situations where the classical EM spectral density can
be used, the EM environment is then a continuum which can be treated as a
bath. This makes all the theoretical tools of the field of open quantum systems
available. In the weak-coupling regime, the bath can be treated perturbatively
through the Markov approximation, which just induces level shifts (often assumed
to be included in the emitter frequency $\omega_0$ and thus neglected) and radiative decay
with rate $\gamma_r = 2\pi J(\omega_0)$.

\begin{figure*}[htbp]
  \includegraphics[width=\linewidth]{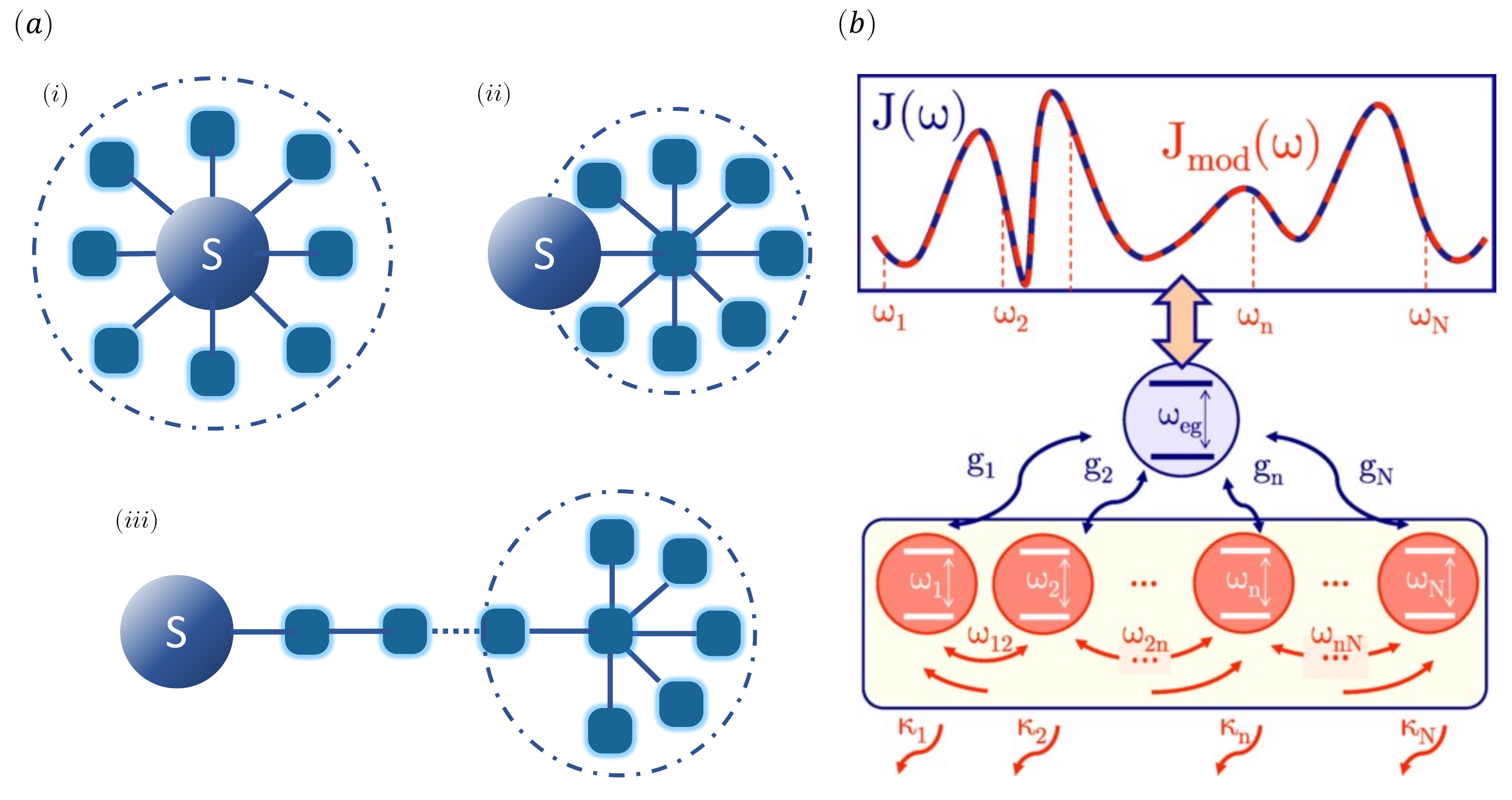}
  \caption{(a): Sketch of the chain mapping model.
  (i) quantum system coupled to a discrete set of environment modes;
  (ii) quantum system coupled to the (collective) reaction mode, with this mode coupled to a
  residual bath of modes; (iii) chain mapping for the environment modes after n
  steps, with a residual bath of N-n modes at the end of the chain.
  (b): Sketch of the few-mode quantization model for one quantum emitter.
  The spectral density of the model can be fitted to the one obtained
  classically from the Green's function. Therefore, the couplings to the
  interacting modes $g_i$, their frequencies and coupling $\omega_{ij}$ and the
  dissipative terms $\kappa_i$ can be known.
  Reproduced with permission.}
  \label{medium&chain}
\end{figure*}

In the more interesting situation where the Markovian approximation is not
applicable, there are several available methods to treat a bath of harmonic
oscillators (the photon modes) exactly using advanced computational tools. One
of these tools is the cumulant expansion~\cite{Kubo1962, Kira2008, Kira2011},
which is based on solving the Heisenberg equations of motion for the
correlations between operators and truncating the resulting expansion at a given
order. It is in this sense an extension of mean-field approaches to arbitrary
order. For small numbers of molecules (or a single one), relatively high orders
of the correlations are needed to accurately describe the dynamics even in the
presence of photonic continua~\cite{Sanchez-Barquilla2020}, while for large
numbers of molecule, the expansion converges much
earlier~\cite{Arnardottir2020}. 

Several other commonly used approaches rely on the so-called chain mapping, an
orthogonal transformation that maps the Hamiltonian of one emitter coupled to a
continuum of modes to a chain-like Hamiltonian where the emitter is only coupled
to the first site, which corresponds to a reaction mode, i.e., a collective
environment mode, in an infinite string of modes coupled through
nearest-neighbor interactions~\cite{Chin2010, Prior2010, Woods2014}.
\autoref{medium&chain}a shows a sketch of this transformation. In this form,
tensor network approaches that represent a high-dimensional wave function as a
product of many lower-dimensional matrices (a so-called matrix product state)
become highly efficient as the entanglement in an effective 1D system such as a
chain is limited. Tensor network approaches depend exactly on a truncation of
the possible entanglement between different parts of the system. In the context
of molecular polaritonics, they have been shown to allow the description of
several molecules coupled to complex environments~\cite{DelPino2018Ground,
DelPino2018Dynamics, *DelPino2018DynamicsErratum, Zhao2020}. A fully converged
tensor network calculation gives exact results, but becomes computationally
challenging when long propagation times are desired as the entanglement grows
over time. Furthermore, the formally infinite chain has to be truncated at
finite length in any realistic calculation, with the required length increasing
with propagation time (to prevent unphysical reflections from the end of the
chain). In order to decrease the length of the chain and access long times with
low computational effort, it is possible to introduce fictitious losses along it
that lead to damping of the propagating excitations (similar to the absorbing
potentials used in many areas of physics)~\cite{Sanchez-Barquilla2021}. Another
approach is to employ transfer tensors, which can be used to propagate to
arbitrary times with linear computational cost~\cite{Rosenbach2016}.

While the description of the EM modes as a structured continuum described by the
spectral density is formally exact, it is often advantageous and desired to
obtain a description of the environment in terms of a few discrete modes,
corresponding to the physical image of isolated cavity modes. When losses are
included, these are not true eigenmodes of the system, but resonances with a
given linewidth embedded in the continuum. Several methods to achieve such a
few-mode quantized description have been developed in the past few years. One is
based on quasinormal modes, which are eigenmodes of the Maxwell equations
including losses with complex frequencies~\cite{Ge2014Quasinormal,Sauvan2021}.
They can be used to expand the electric field solutions based on a master
equation approach~\cite{Yang2015Analytical}, or explicitly quantized such that
the EM field is represented in terms of discrete bosonic
modes~\cite{Franke2019}. An alternative approach that does not require
calculation and explicit quantization of quasinormal modes is based on the fact
that two systems with the same spectral density are indistinguishable for an
emitter. This allows the construction of a model system consisting of a few
coupled discrete modes that are themselves coupled to a background bath and
reproduce the full spectral density~\cite{Medina2021}. \autoref{medium&chain}b
shows a sketch of that model. The parameters of the model, which are obtained
through fitting of the spectral density, are the coupling between the emitter
and the discrete modes $g_i$, the frequencies of these modes and their couplings
$\omega_{ij}$, and their dissipation $\kappa_i$. The coupling to the background
modes is spectrally flat (by construction) and thus leads to Markovian dynamics
that can be represented in a Lindblad master equation, such that the full EM
continuum is represented by a collection of lossy and coupled discrete modes in
the master equation. This approach is not only computationally efficient, but
also allows to describe the EM environment in the language of cavity QED and
quantum optics. The fact that the discrete modes are mutually coupled makes it
able to reproduce even complex interference phenomena between the EM modes in
the spectral density. While originally developed for a single
emitter~\cite{Medina2021}, we have recently extended the approach to the case of
several emitters~\cite{Sanchez-Barquilla2021Few-Mode}, or one emitter with
several contributing transitions (such as different orientations of the dipole
moments). This is enabled by defining a generalized spectral density that fully
encodes the interaction between the EM modes and several emitters,
\begin{equation}
  \mathcal{J}_{nm}(\omega) = \frac{\hbar\omega^2}{\pi\epsilon_0 c^2} \mathbf{n}_n\cdot \Im\mathbf{G}(\rr_n,\rr_m,\omega)\cdot \mathbf{n}_m,
\end{equation}
and which is again determined by the classical dyadic Green's functions. Here,
$\mathbf{n}_n$ is the unit vector describing the orientation of the dipole
transitions that are taken into account, while $\rr_n$ is the position of the
corresponding emitter. Note that in this definition, the dipole moment is not
included in the spectral density as emitters with many levels and dipole
transitions between them can be treated.

\section{Introducing molecular complexity}

For the description of the molecules, we focus on approaches that are
well-adapted for describing ``good'' molecular emitters, meaning ones where the
first electronically excited state is relatively stable against nonradiative
decay and photochemical reactions do not take place. For such emitters, it is
often a good approximation to represent the nuclear potential energy surfaces as
harmonic oscillators, which significantly simplifies the treatment. When more
chemical detail is needed, a wide variety of methods are nowadays available, but
doing so typically limits the level of description of nuclear
motion~\cite{Fregoni2021Perspective}.

As detailed in the introduction, the simplest approach is to treat molecules as
two-level systems. This can be well-justified for studies at cryogenic
temperatures where vibrational sublevels are individually resolved and
addressable~\cite{Wang2019Turning, Pscherer2021}. At room temperature, the
influence of the vibrational motion of the molecule can be approximately
included by adding a pure dephasing term in a Lindblad master equation
description. This can be understood as arising from an exciton-vibration
coupling term treated through the Markov approximation. However, since under
strong coupling the molecular exciton state gets distributed over the
polaritonic modes, the Markov approximation that was originally performed for
the emitter by itself to obtain a pure dephasing term is not valid
anymore~\cite{Carmichael1973}. Including it without further modification in the
strongly coupled system leads to artificial population transfer between the
polaritons and dark states, with equal rates for pumping of energy from the
reservoir of vibrations to the system as for loss of energy from the system to
the reservoir. This is unphysical when the Rabi splitting is comparable to or
larger than the thermal energy, a condition that is essentially always fulfilled
in molecular exciton-polariton strong coupling. This problem can be resolved by
applying the Markov approximation after taking into account the strong coupling,
e.g., by using a Bloch-Redfield approach~\cite{DelPino2015Quantum,
Saez-Blazquez2018Organic, Saez-Blazquez2019}.

When exciton-phonon coupling is sufficiently strong that the above approach
breaks down, it becomes necessary to explicitly include the vibrational modes of
the molecules. The simplest approach is the so-called Holstein model, which
treats only a single vibrational mode and describes each molecule as two
displaced harmonic oscillators. For multiple emitters, this leads to the
so-called Holstein-Tavis-Cummings model~\cite{Kirton2013, Herrera2016}, within
which each electronic level is represented by several vibrational sub-levels.
Within this model, the effect of the vibronic coupling can be studied, so an
analysis of nuclear dynamics in molecules under strong coupling is feasible.
Along this line, it has been predicted that electron transfer between different
excited states can be enhanced or suppressed~\cite{Herrera2016}. The inclusion
of vibronic sublevels and the concomitant emergence of dark vibronic polaritons
(collective light-matter states that weakly absorb but strongly emit radiation)
allows for a better description of the spectroscopy and dynamics of organic
microcavities in the strong coupling regime~\cite{Herrera2017Dark,
Herrera2017Absorption, Herrera2018Theory}. The inclusion of vibrational levels
is also highly relevant for the description of phenomena such as organic polariton
lasing and polariton condensation~\cite{Kirton2013, Zeb2018, Strashko2018}.

An extension of the Holstein-Tavis-Cummings model that allows for more realistic
molecular structure is to consider more complex potential energy curves instead
of harmonic oscillators, permitting the treatment of vibrational nonlinearities
and (photo)chemical reactions within a relatively simple model (especially if
only a single vibrational degree of freedom is treated). The hybridization of
the potential energy surfaces in the strong coupling regime then leads to
hybridized polaritonic potential energy surfaces (PoPES), which have mixed
photon-matter properties~\cite{Galego2015, Galego2016, Feist2018}. A schematic
representation of these PoPES can be found in the right bottom panel of
\autoref{fig:scheme}.

Assuming that vibrational modes are harmonic oscillators but taking into account
all degrees of freedom for the molecules (typically hundreds), and potentially
of the surrounding solvent or polymer host material, one can again rely on
tensor network techniques as discussed above for the photonic modes. The
molecular vibrations are then represented by an independent chain of harmonic
oscillators~\cite{DelPino2018Ground, DelPino2018Dynamics,
*DelPino2018DynamicsErratum, Zhao2020}. Alternatively, the bath of harmonic
vibrational modes can be represented through its correlation function and
simulated using the time-evolving matrix product operator (TEMPO)
method~\cite{Strathearn2018, Fux2021}. This uses a tensor network to describe
the system's history over a finite memory time and can thus represent
non-Markovian dynamics. Combining this technique with a mean-field approximation
further reduces the problem size~\cite{Fowler-Wright2021}.

We note at this point that although many effects can be understood by the use of
the previous models, all of them constitute strong approximations for the
molecular structure. In particular, the use of harmonic oscillators to describe
the vibrational modes precludes the description of any nonlinear vibrational
effects or of chemical transition states, conical intersections, etc. However, a
full quantum description of the molecules is an extremely challenging task and
only possible for small molecules. To give a more complete picture by including
all degrees of freedom inside the molecules requires utilizing quantum chemistry
and ab-initio approaches~\cite{Fregoni2021Perspective}.

\section{Summary and outlook}

In this article we have provided a perspective on the current status of the
theoretical investigation devoted to analyzing the exciting physics in the
emergent field of molecular polaritonics. This area of research deals with the
strong light-matter coupling regime that appears between electronic/vibrational
excitations within (organic) molecules and confined light fields. As for the
light field component, a photonic structure that acts as a cavity is needed.
Depending on the cavity used, strong coupling can be reached by utilizing a
large ensemble of molecules or just one or a few of them, depending on whether
or not subwavelength confinement is achieved. 

The theoretical description of the EM modes that arise in these photonic
structures, which in general are lossy, is then a challenging task. When the
cavities present small losses, they can be treated perturbatively, while for
small enough subwavelength cavities, the quasistatic approximation can be used,
which allows for semi-analytical solutions. However, in many physical
situations, a fully quantized description of the EM field in photonic structures
is required. Here we have shown how the macroscopic QED formalism provides the
necessary theoretical and numerical tools to accurately describe light-matter
strong coupling in arbitrary structures. Within this framework, the spectral
densities that characterize the coupling between one or several quantum emitters
and confined EM modes are fully determined by the classical Green's functions,
which are calculated using standard numerical solvers of Maxwell's equations in
complex EM media. Based on macroscopic QED, it is then feasible to develop
approaches that allows for a tractable treatment of complex photonic
environments. Among these simplified treatments, the most promising approaches
are those based on the concept of quasinormal EM modes and a very recent one
that relies on the construction of a model system involving a small number of
lossy and interacting EM modes whose parameters are fitted to exactly reproduce
the spectral densities associated with the photonic structure under study.

Regarding the matter component, organic molecules also have a complex internal
structure, which prevent them from being theoretically modelled as just
two-level systems in many situations. To include molecular complexity in the
theoretical description, it is then mandatory to add some ingredients to the
standard two-level model to describe the vibrational modes of the molecules.
Depending on the strength of the vibronic coupling, a pure dephasing term, the
Bloch-Redfield approximation or the explicit inclusion of the vibrational modes
need to be utilized. Within this last approach, the most used framework is the
so-called Holstein-Tavis-Cummings model, which only takes into account one
vibrational mode, modelled as a harmonic oscillator, which has proven to be very
successful in providing physical insight. Going beyond this model can be
achieved either by including more of the (typically hundreds of) vibrational
modes of a molecule, or by substituting the harmonic oscillators by more
realistic potential energy surfaces. The first approach captures
vibration-induced dephasing and decoherence, while the second naturally accounts
for the vibrational nonlinearities and has also allowed for a fundamental
description of (photo)chemical reactions induced by strong light-matter
coupling. Nevertheless, in order to have a more accurate description of the
internal structure of organic molecules, numerical formalisms that rely on
quantum chemistry codes or ab initio approaches need to be utilized.

At this stage, for the case of collective strong coupling in which a large
(macroscopic) number of molecules is involved, the current status of the
theoretical research on molecular polaritonics does not allow for both a
realistic treatment of the internal structure of the organic molecules and a
detailed account of the complex EM media. This is indeed very frustrating as the
majority of the experiments carried out in this field belong to the category of
collective strong coupling. Therefore, a quantitative agreement between
ab-initio theory and experiment is not within reach nowadays. This is why during
the last decade, most of theoretical research has focused either on giving
fundamental support to some of the experimental findings or to propose new
effects that result from theoretical approaches based on simplified models. On a
more positive note, for the case in which only a single or few molecules
participate in the strong coupling phenomenon, after ten years of intense
research, we now have the adequate theoretical and numerical tools to accurately
describe both the internal vibrational modes of the organic molecules and the
complexity of the sub-wavelength EM fields associated with nanophotonic (mainly
plasmonic) structures. Then, we expect that all this theoretical knowledge will
help to open new and exciting avenues for research in molecular polaritonics.

This work has been funded by the European Research Council through Grant
ERC-2016-StG-714870 and by the Spanish Ministry for Science, Innovation, and
Universities -- Agencia Estatal de Investigación through grants
RTI2018-099737-B-I00, PCI2018-093145 (through the QuantERA program of the
European Commission), and CEX2018-000805-M (through the María de Maeztu program
for Units of Excellence in R\&D). We also acknowledge financial support from the
Proyecto Sinérgico CAM 2020 Y2020/TCS-6545 (NanoQuCo-CM) of the Community of
Madrid.

\bibliography{references}

\end{document}